\documentstyle[12pt,epsfig]{elsart}
\begin{document}

\newcommand{\re}{\mathop{\mathrm{Re}}}
\newcommand{\im}{\mathop{\mathrm{Im}}}
\newcommand{\D}{\mathop{\mathrm{d}}}
\newcommand{\I}{\mathop{\mathrm{i}}}

\noindent {\Large DESY 04-045}

\noindent {\Large March 2004}

\begin{frontmatter}

\journal{Optics Communications}

\title{A new technique to generate 100 GW-level attosecond X-ray
pulses from the X-ray SASE FELs}

\author{E.L.~Saldin},
\author{E.A.~Schneidmiller},
and \author{M.V.~Yurkov}

\address{Deutsches Elektronen-Synchrotron (DESY),
Notkestrasse 85, Hamburg, Germany}

\begin{abstract}

We propose a scheme for generation of single 100 GW 300-as pulse in the
X-ray free electron laser with the use of a few cycles optical pulse
from Ti:sapphire laser system. Femtosecond optical pulse interacts with
the electron beam in the two-period undulator resonant to 800 nm
wavelength and produces energy modulation within a slice of the
electron bunch. Following the energy modulator the electron beam enters
the first part of the baseline gap-adjustable X-ray undulator and
produces SASE radiation with 100 MW-level power. Due to energy
modulation the frequency is correlated to the longitudinal position
within the few-cycle-driven slice of the SASE radiation pulse. The
largest frequency offset corresponds to a single-spike pulse in the
time domain which is confined to one half-oscillation period near the
central peak electron energy. After the first undulator the electron
beam is guided through a magnetic delay which we use to position the
X-ray spike with the largest frequency offset at the "fresh" part of
the electron bunch.  After the chicane the electron beam and the
radiation produced in the first undulator enter the second undulator
which is resonant with the offset frequency.  In the second undulator
the seed radiation at reference frequency plays no role, and only a
single (300~as duration) spike grows rapidly. The final part of the
undulator is a tapered section allowing to achieve maximum output power
100-150 GW in 0.15 nm wavelength range. Attosecond X-ray pulse is
naturally synchronized with its fs optical pulse which reveals unique
perspective for pump-probe experiments with sub-femtosecond resolution.

\end{abstract}

\end{frontmatter}

\clearpage

\thispagestyle{empty}

$\mbox{}$

\clearpage

\setcounter{page}{1}

\section{Introduction}

Time-resolved experiments are used to monitor time-dependent phenomena.
The study of dynamics in physical systems often requires time
resolution beyond the femtosecond capabilities. Subfemtosecond
capabilities are now available in the XUV wavelength range
\cite{laser-atto1,laser-atto2}. This is achieved by focusing a fs laser
into a gas target creating radiation of high harmonics of fundamental
laser frequency. In principle, table-top ultra-fast X-ray sources have
the right duration to provide us with a view of subatomic
transformation processes. However, their power and photon energy are by
far low. There also exists a wide interest in the extension
of attosecond techniques into the 0.1 nm wavelength range.

With the realization of the fourth-generation light sources operating in
the X-ray regime \cite{xfel-tdr,lcls-cdr}, new attoscience experiments
will become possible. In its initial configuration the XFEL pulse
duration is about 100 fs, which is too long to be sufficient for this
class of experiments. The generation of subfemtosecond X-ray pulses is
critical to exploring the ultrafast science at the XFELs.  The advent
of attosecond X-ray pulses will open a new field of time-resolved
studies with unprecedented resolution. X-ray SASE FEL holds a great
promise as a source of radiation for generating high power, single
attosecond pulses. Recently a scheme to achieve pulse duration down to
attosecond time scale at the wavelengths around 0.1 nm has been
proposed \cite{atto-1}. It has been shown that by using X-ray SASE FEL
combined with terawatt-level, sub-10-fs Ti:sapphire laser system it
will be possible to produce GW-level X-ray pulses that are reaching 300
attoseconds in duration. In this scheme an ultrashort laser pulse is
used to modulate the energy of electrons within the femtosecond slice
of the electron bunch at laser frequency. Energy-position correlation in
the electron pulse results in spectrum-position correlation in the SASE
radiation pulse. Selection of ultra-short X-ray pulses is achieved by
using the monochromator. Such a scheme for production of single
attosecond X-ray pulses would offer the possibility for pump-probe
experiments, since it provides a precise, known and tunable interval
between the laser and X-ray sources.

In this paper we propose a new method allowing to increase output power
of attosecond X-ray pulses by two orders of magnitude. It is based on
application of sub-10-fs laser for slice energy modulation of the
electron beam, and application "fresh bunch" techniques for selection
of single attosecond pulses with 100 GW-level output power. The
combination of very high peak power (100 GW) and very short pulse (300
as) will open a vast new range of applications. In particular, we
propose visible pump/X-ray probe technique that would allow time
resolution down to subfemtosecond capabilities. Proposed technique
allows to produce intense ultrashort X-ray pulses directly from the
XFEL, and with tight synchronization to the sample excitation laser.
Another advantage of the proposed scheme is the possibility to remove
the monochromator (and other X-ray optical elements) between the X-ray
undulator and a sample and thus to directly use the probe attosecond
X-ray pulse.

\section{High power attosecond facility description}

\begin{figure}[b]
\begin{center}
\epsfig{file=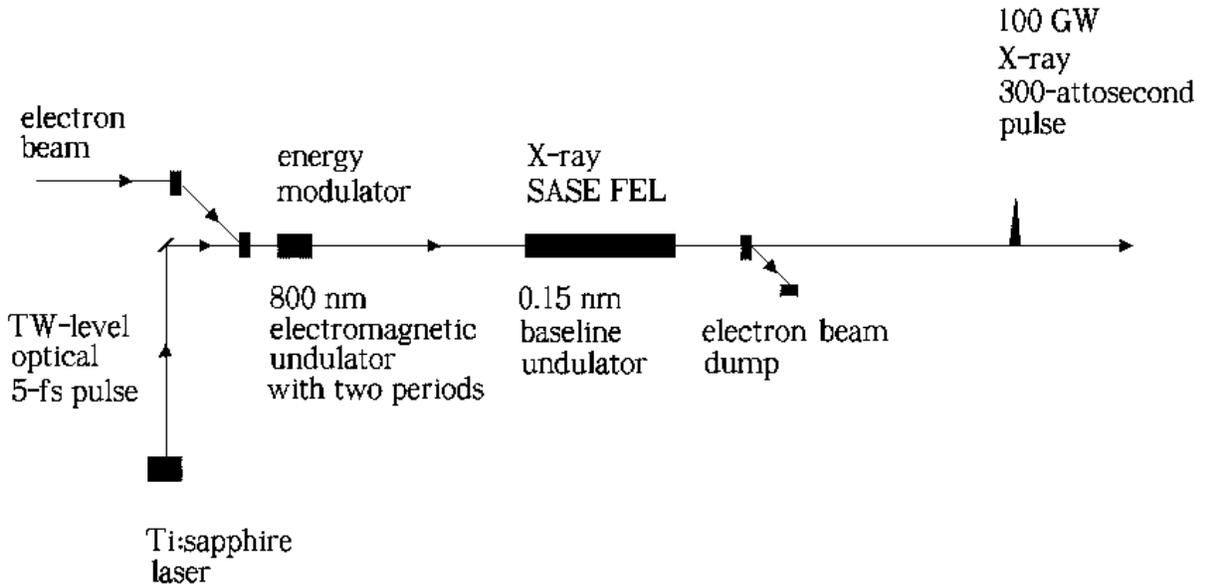,width=1\textwidth}
\end{center}
\caption{Schematic diagram of high power attosecond X-ray source}
\label{fig:x4}
\end{figure}

A basic scheme of the high-power attosecond X-ray source is shown in
Fig. \ref{fig:x4}. An ultrashort laser pulse is used to modulate the
energy of electrons within the femtosecond slice of the electron bunch
at laser frequency. The seed laser pulse will be timed to overlap with
the central area of the electron bunch.  It serves as a seed for
modulator which consists of a short (a few periods) undulator.
Following the energy modulator the beam enters the baseline
(gap-tunable) X-ray undulator.  In its simplest configuration the X-ray
undulator consists of an uniform input undulator and  nonuniform
(tapered) output undulator separated by a magnetic chicane (delay). The
process of amplification of radiation in the input undulator develops
in the same way as in conventional X-ray SASE FEL:  fluctuations of the
electron beam current serve as the input signal \cite{book}. When an
electron beam traverses an undulator, it emits radiation at the
resonance wavelength $\lambda = \lambda_{\mathrm{w}}(1 +
K^{2}/2)/(2\gamma^{2})$.  Here $\lambda_{\mathrm{w}}$ is the undulator
period, $mc^{2}\gamma$ is the electron beam energy, and $K$ is the
undulator parameter. In the proposed scheme the laser-driven sinusoidal
energy chirp produces a correlated frequency chirp of the resonant
radiation $\delta\omega/\omega \simeq 2\delta\gamma/\gamma$.

\begin{figure}[tb]
\begin{center}
\epsfig{file=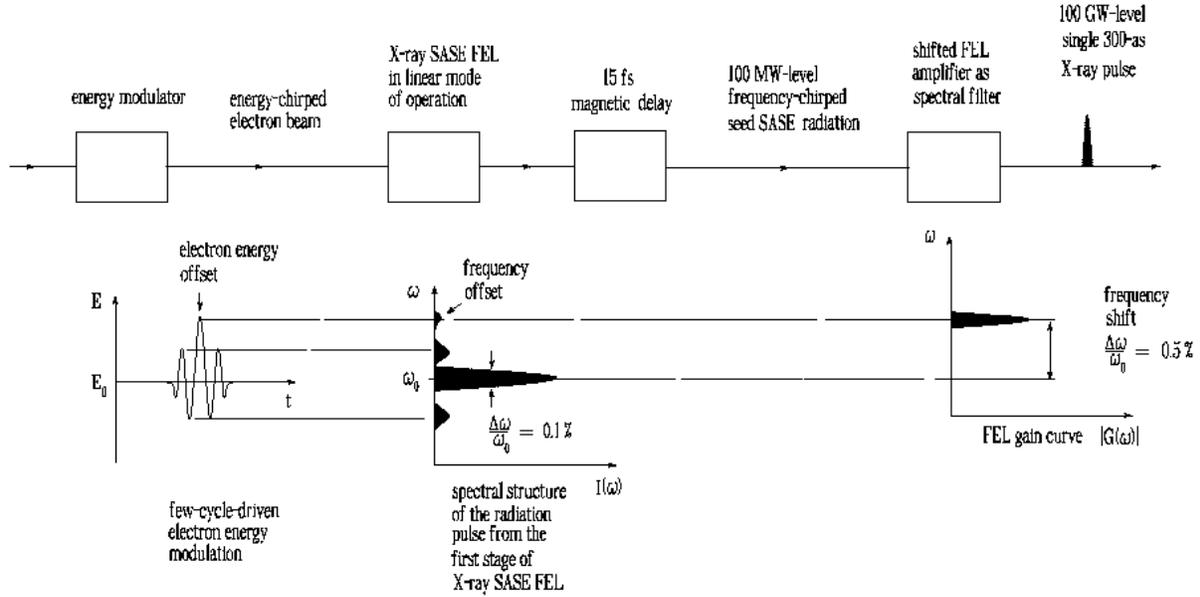,width=1\textwidth}
\end{center}
\caption{Sketch of high power attosecond X-ray pulse synthezation
through frequency chirping and spectral filtering }
\label{fig:x5}
\end{figure}

\begin{figure}[tb]
\begin{center}
\epsfig{file=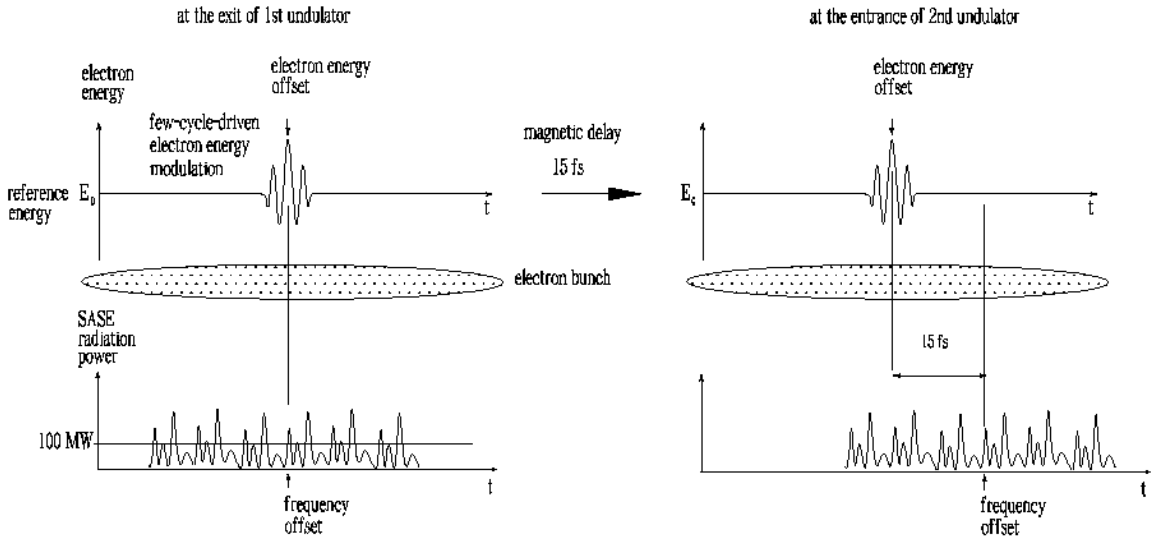,width=1\textwidth}
\end{center}
\caption{Sketch of principle of "fresh bunch" technique}
\label{fig:x7}
\end{figure}

Our concept of attosecond X-ray facility is based on the use of a few
cycle optical pulse from Ti:sapphire laser system. This optical pulse
is used for modulation of the energy of the electrons within a slice of
the electron bunch at a wavelength of 800 nm. Due to extreme temporal
confinement, moderate optical pulse energies of the order of a few mJ
can result in electron energy modulation amplitude higher than 30-40
MeV. In few-cycle laser fields high intensities can be "switched on"
nonadiabatically within a few optical periods. As a result, a central
peak electron energy modulation is larger than other peaks. This
relative energy difference is used for selection of SASE radiation
pulses with a single spike in time domain. Single-spike selection can
effectively be achieved when electron bunch passes through a magnetic
delay and output undulator operating at a shifted frequency. A
schematic, illustrating these processes, is shown in Figs.~\ref{fig:x5}
and \ref{fig:x7}.

\begin{figure}[tb]
\begin{center}
\epsfig{file=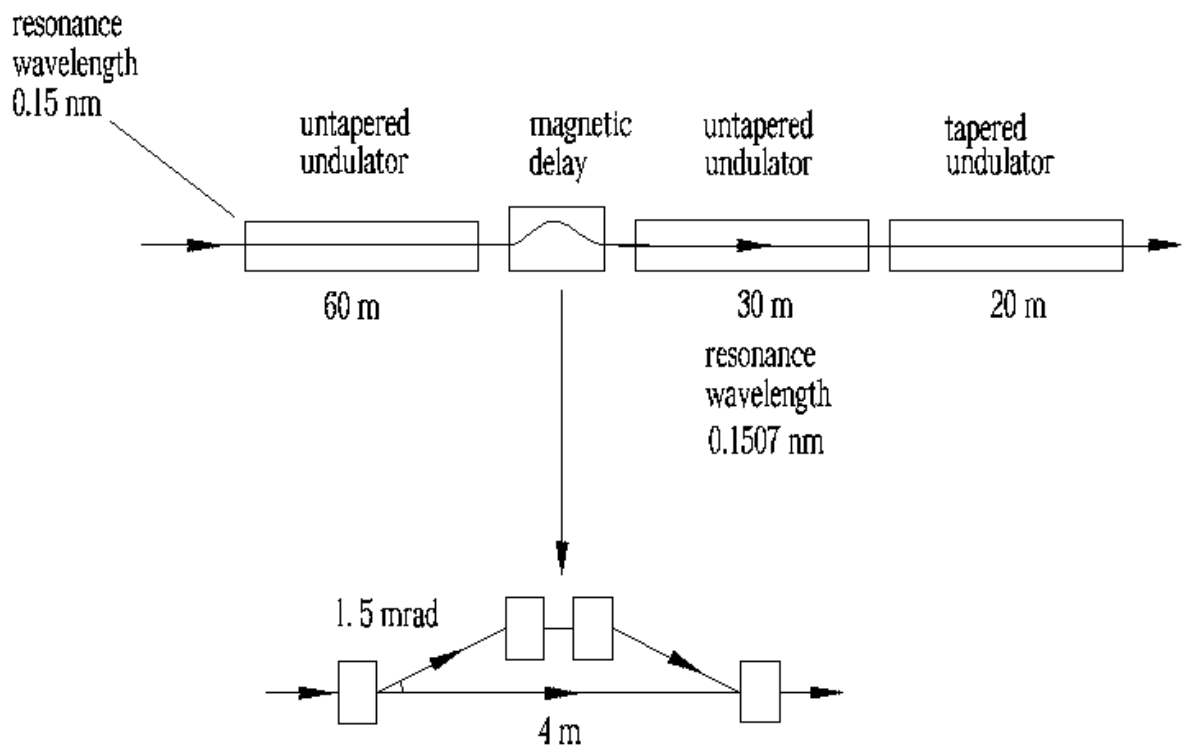,width=1\textwidth}
\end{center}
\caption{Design of undulator system for high power attosecond X-ray
source
}
\label{fig:x3}
\end{figure}

The input undulator is a conventional 0.15 nm SASE FEL operating in the
high-gain linear regime. This undulator is long enough (60 m) to reach
100 MW-level output power (see Fig. \ref{fig:x3}).  After the input
undulator the electron beam is guided through a magnetic delay
(chicane). The trajectory of the electron beam in the chicane has the
shape of an isosceles triangle with the base equal to $L$. The angle
adjacent to the base, $\theta$, is considered to be small.  Parameters
in our case are: $L = 4$ m, $\theta = 1.5$ mrad, compaction factor
$L\theta^{2} = 8 \mu$m, extra path length $L\theta^{2}/2 = 4 \mu$m,
horizontal offset $L\theta/2 = 3$ mm.  In the present design we have
only 4 $\mu$m extra path length for the electron beam, while the FWHM
length of electron bunch is about 50 $\mu$m. Calculations of the
coherent synchrotron radiation effects show that this should not be a
serious limitation in our case.

Passing the chicane the electron beam and seed SASE radiation enter
the output undulator operating at an offset frequency. We use a magnetic
delay to position the offset frequency radiation at the "fresh" part of
the electron bunch.  This seed single spike at an offset frequency starts
interacting with the new set of electrons, which have no energy
modulation, since they did not participate in the previous interaction
with optical laser pulse (see Fig. \ref{fig:x7}). This is
the essence of the "fresh bunch" techniques which was introduced in
\cite{yu-freshb}.

\begin{figure}[tb]
\begin{center}
\epsfig{file=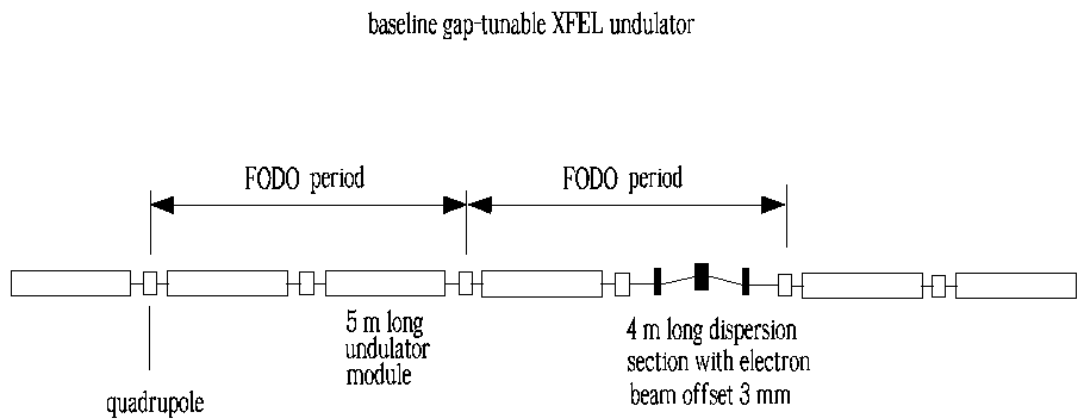,width=0.9\textwidth}
\end{center}
\caption{Installation of a magnetic delay in the baseline XFEL
undulator. The quadrupole separation of a FODO lattice is large enough
so that magnetic chicane of length 4 m can be installed }
\label{fig:x2}
\end{figure}

\begin{figure}[tb]
\begin{center}
\epsfig{file=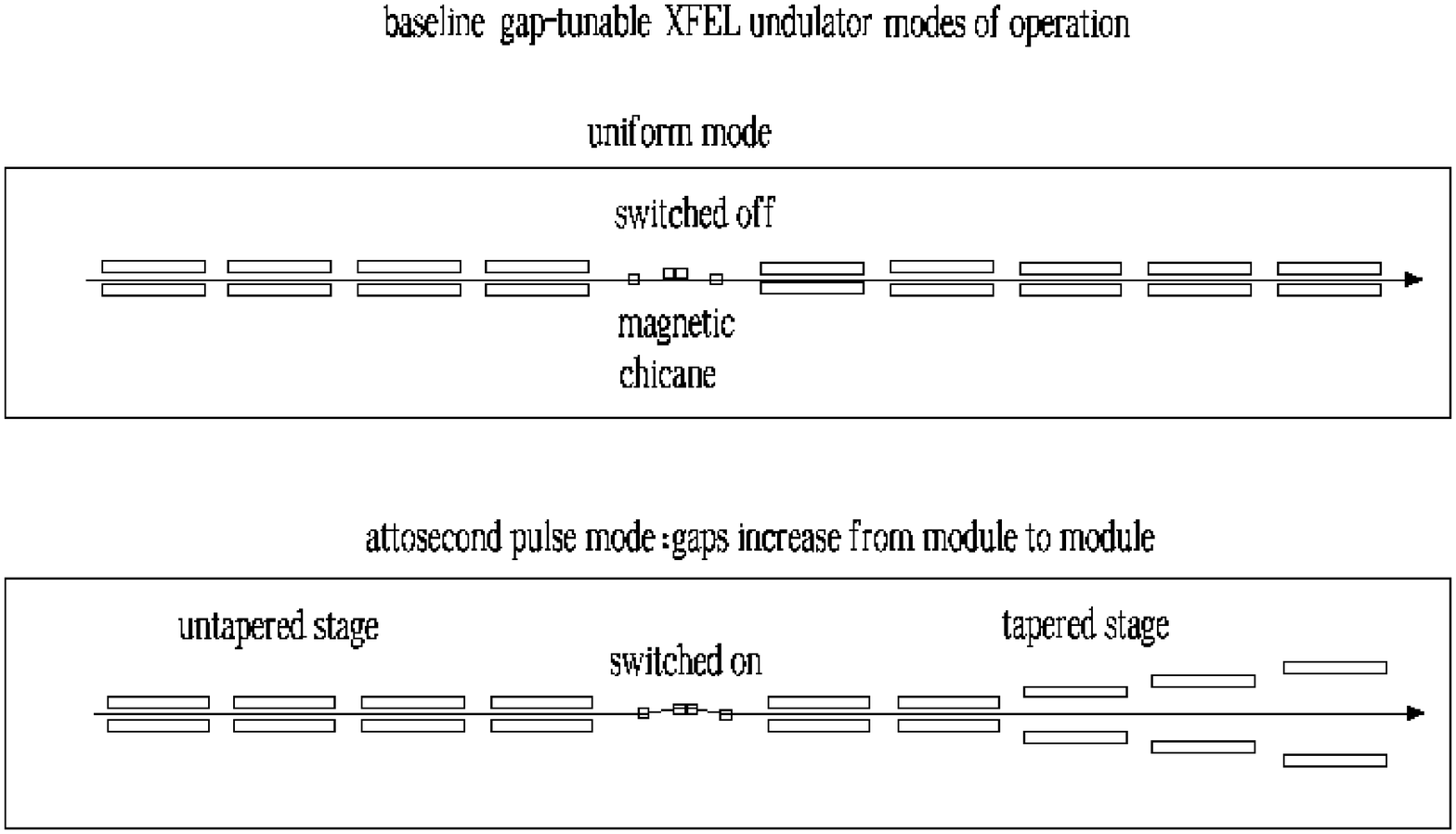,width=1\textwidth}
\end{center}
\caption{Baseline XFEL undulator allowing different modes of operation
for SASE FEL. Tuning of the radiation wavelength and magnetic field
tapering is provided by changing the gap}
\label{fig:x1}
\end{figure}

In the output undulator seed radiation at reference frequency plays no
role. However, single spike at an offset frequency is exponentially
amplified upon passing through the first (uniform) part of the output
undulator. This part is long enough  (30 m) to reach saturation. The
power level at saturation is about 20 GW. The most
promising way to extend output power is the method of tapering the
magnetic field of the undulator.  Tapering consists in slowly reducing
the field strength of the undulator field to preserve the resonance
wavelength as the kinetic energy of the electrons changes.
The strong radiation field produces a ponderomotive well which is deep
enough to trap the particles. The radiation produced by these captured
particles increases the depth of the ponderomotive well, and they are
effectively decelerated. As a result, much higher power can be achieved
than for the case of a uniform undulator. At the total tapered
undulator length of 20 m, the single-spike power is enhanced by a
factor of five, from 20 GW-level  to 100 GW-level. With the baseline
gap-tunable undulator design this option would require only
installation of a magnetic delay. The quadrupole separation of an
undulator FODO lattice (5 m) is large enough so that relatively short
(4 m) magnetic chicane can be installed  (see Fig. \ref{fig:x2}). The
undulator taper could be simply implemented as a step taper from one
undulator segment to the next.  Figure \ref{fig:x1} shows the design
principle of undulator for attosecond mode of operation.

Our study shows that the method proposed in this paper allows direct
production from XFEL of single 100 GW-level X-ray pulses with FWHM
duration of 300 as. Contrast of a single spike is mainly defined by the
amplification of shot noise in the main part of electron bunch. This
effect leads to degradation of contrast of output attosecond X-ray
pulses. However, parameters of the undulator system can be optimized in
such a way that the contrast degradation is insignificant.  Calculation
shows that in optimum case the ratio of the attosecond pulse power
(signal) and SASE pulse power (background) at the undulator exit
reaches a value of 400.

\section{Generation of 100 GW-level attosecond pulses from XFEL}

\begin{figure}[b]

\epsfig{file=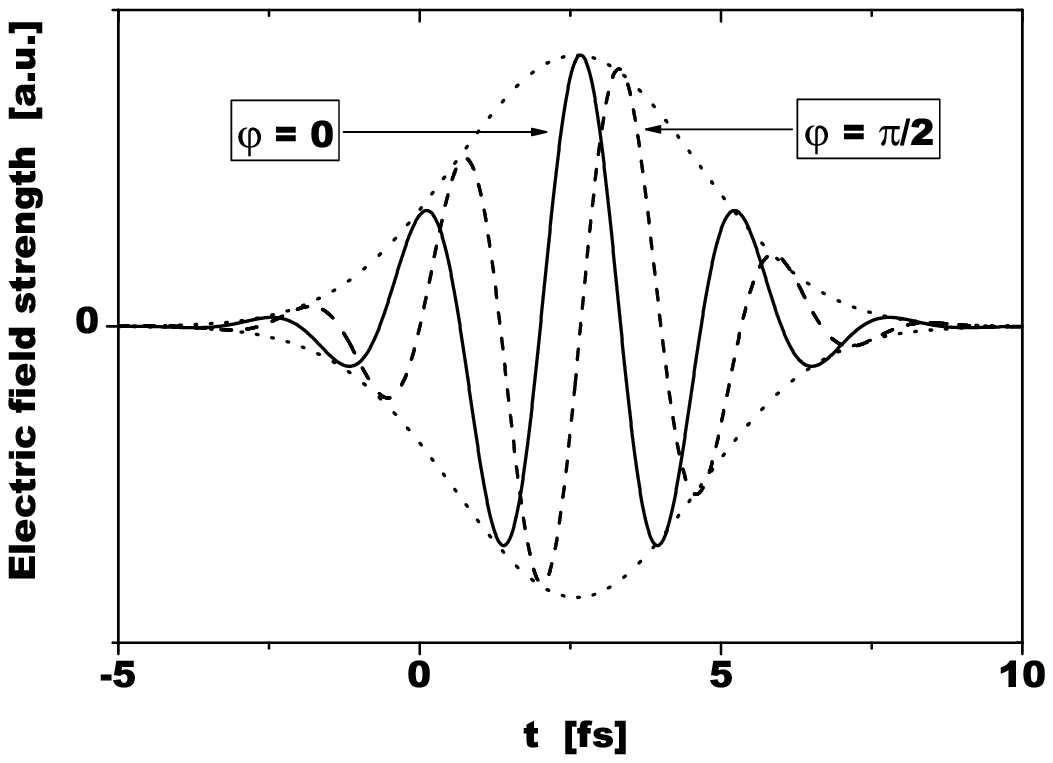,width=0.5\textwidth}

\vspace*{-61mm}

\hspace*{0.5\textwidth}
\epsfig{file=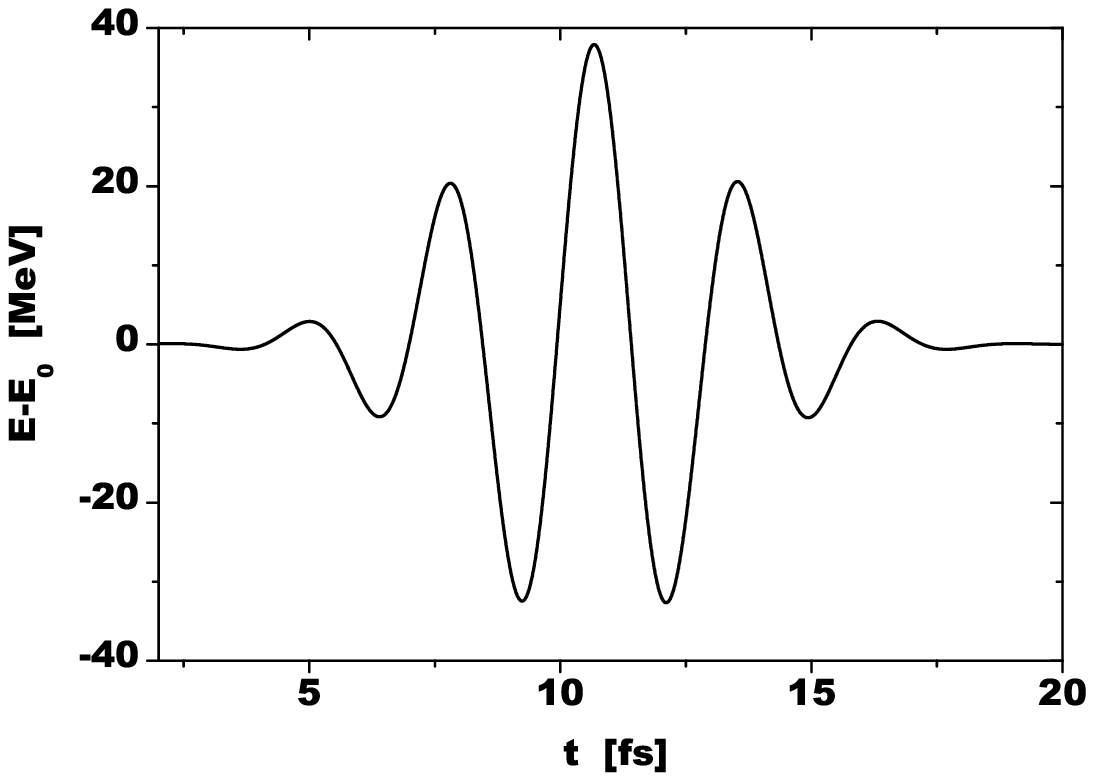,width=0.5\textwidth}

\caption{
Left plot: electric field strength within femtosecond laser pulse.
Right plot: energy modulation of electron bunch at the exit of the
modulator
}
\label{fig:lfield40}
\end{figure}

Operation of 100~GW attosecond SASE FEL is illustrated for the
parameters close to those of the European XFEL operating at the
wavelength 0.15 nm \cite{xfel-tdr}.
Optimization of the attosecond SASE FEL has been performed
with the three-dimensional, time dependent code FAST \cite{fast} taking
into account all physical effects influencing the SASE FEL operation
(diffraction effects, energy spread, emittance, slippage effect, etc.).
The parameters of the electron beam
are: energy 15~GeV, charge 1~nC, rms pulse length 25~$\mu $m, rms
normalized emittance 1.4~mm-mrad, rms energy spread 1~MeV. Undulator
period is 3.65~cm.

The parameters of the seed laser are: wavelength 800 nm, energy in the
laser pulse 2--4 mJ, and FWHM pulse duration 5 fs (see Fig.
\ref{fig:lfield40}). The laser beam is focused onto the electron beam
in a short undulator resonant at the optical wavelength of 800~nm.
Optimal conditions of the focusing correspond to the positioning of the
laser beam waist in the center of the modulator undulator. It is
assumed that the phase of laser field corresponds to "cosine" mode
(solid line with $ \varphi = 0$, see Fig.~\ref{fig:lfield40}).
Parameters of the modulator undulator are: period length 50~cm, peak
field 1.6~T, number of periods 2. The interaction with the laser light
in the undulator produces a time dependent electron energy modulation
as it is shown in Fig.~\ref{fig:lfield40}.  For the laser (FWHM) pulse
duration of 5 fs at a laser pulse energy 2-4 mJ, we expect a central
peak energy modulation of 30-40 MeV.

SASE undulator consists of three sections. First section is 57~meter
long uniform undulator. Second section is 28~meters long uniform
undulator with different resonance frequency. Frequency detuning between
the second and the first undulator section is $\Delta \omega/\omega =
0.6$\%. The third section is 20~meters long tapered undulator. A
dispersion section with 8~$\mu $m net compaction factor is placed
between the first and the second undulator sections.

\begin{figure}[tb]

\begin{center}
\epsfig{file=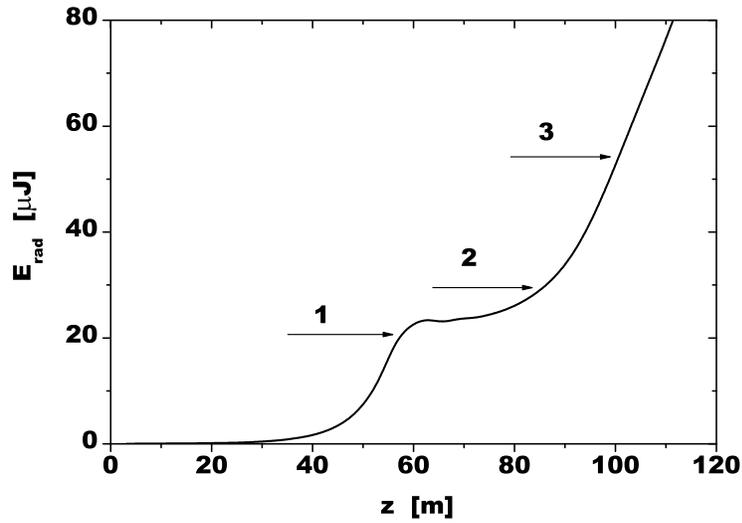,width=0.7\textwidth}
\end{center}

\caption{
Energy in the radiation pulse versus undulator length.
Marks 1, 2, and 3 show the end points of the 1st, 2nd,
and 3rd undulator sections, respectively
}
\label{fig:pz2023}
\end{figure}

\begin{figure}[tb]

\begin{center}
\epsfig{file=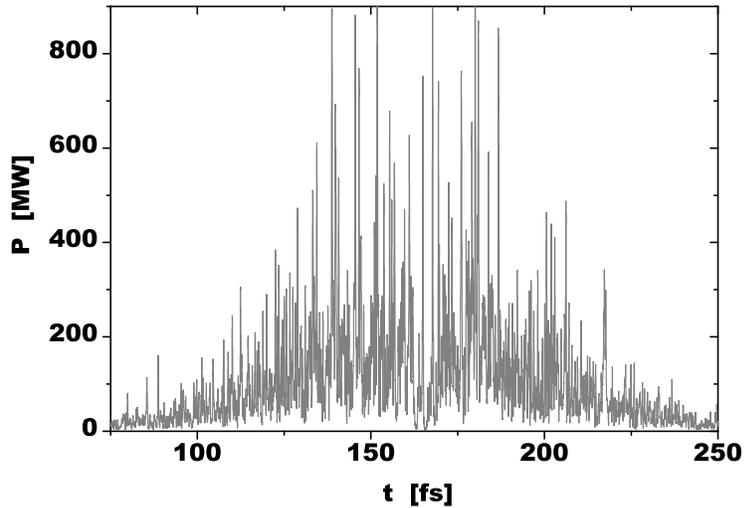,width=0.7\textwidth}
\end{center}

\caption{
Temporal structure of the radiation pulse in the end of the first
undulator section
}
\label{fig:m2318}
\end{figure}

\begin{figure}[tb]

\epsfig{file=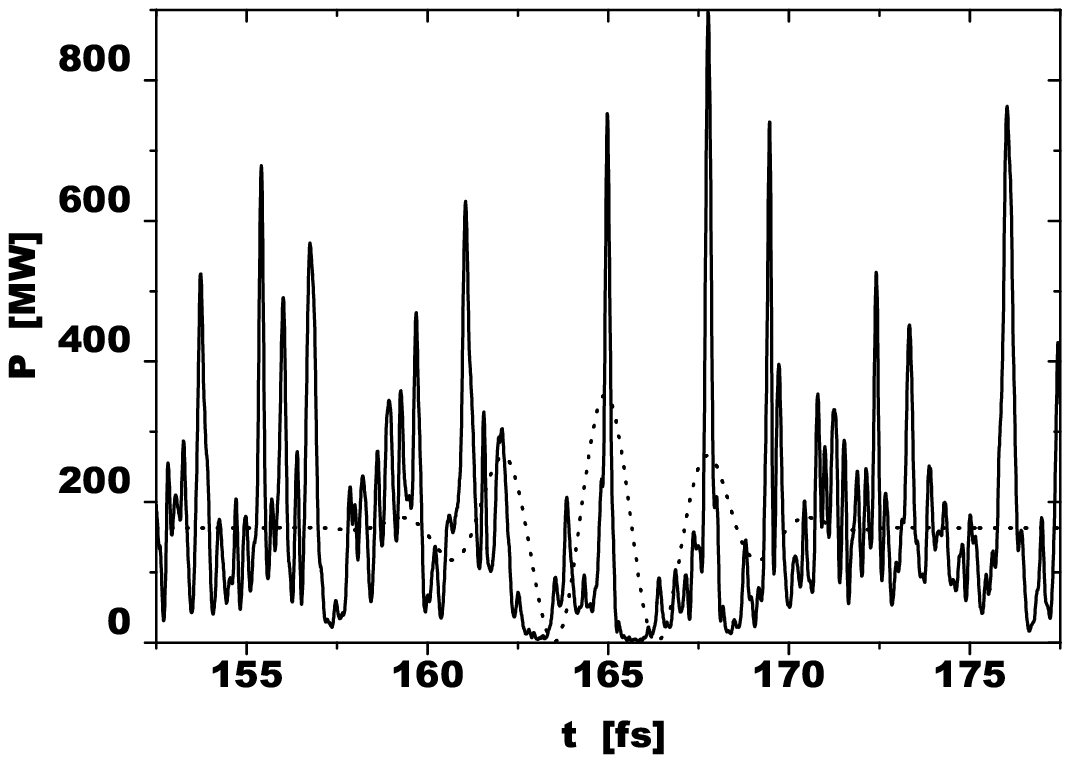,width=0.5\textwidth}

\vspace*{-61mm}

\hspace*{0.5\textwidth}
\epsfig{file=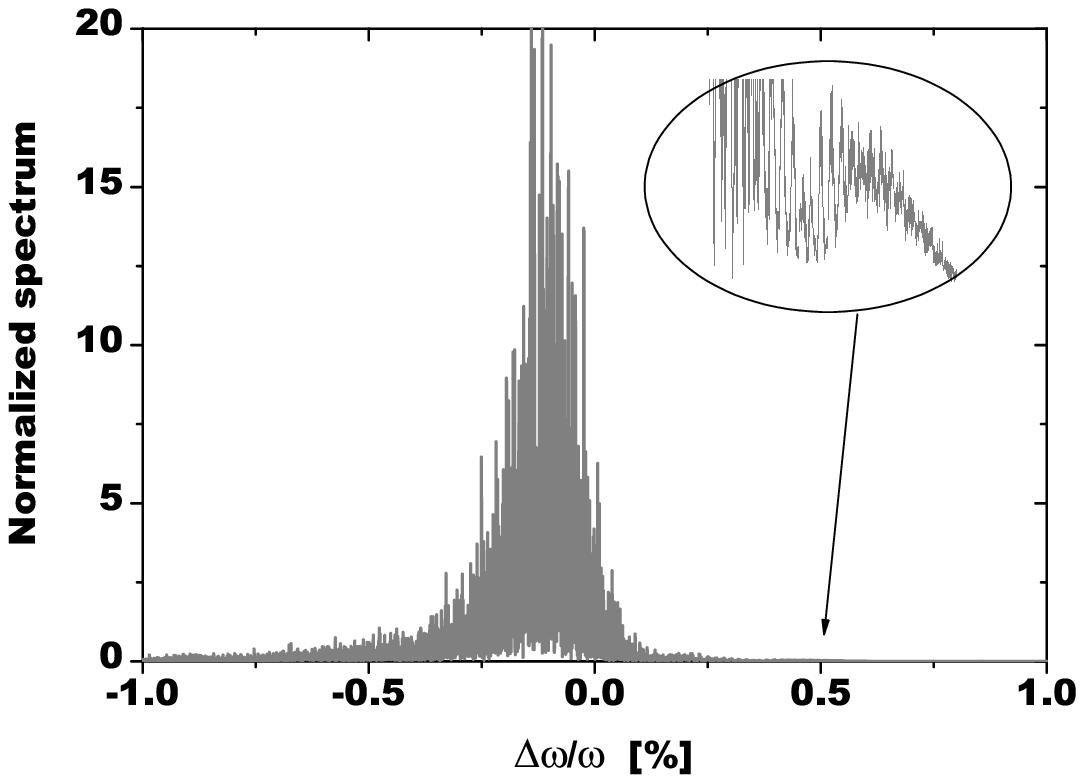,width=0.5\textwidth}

\epsfig{file=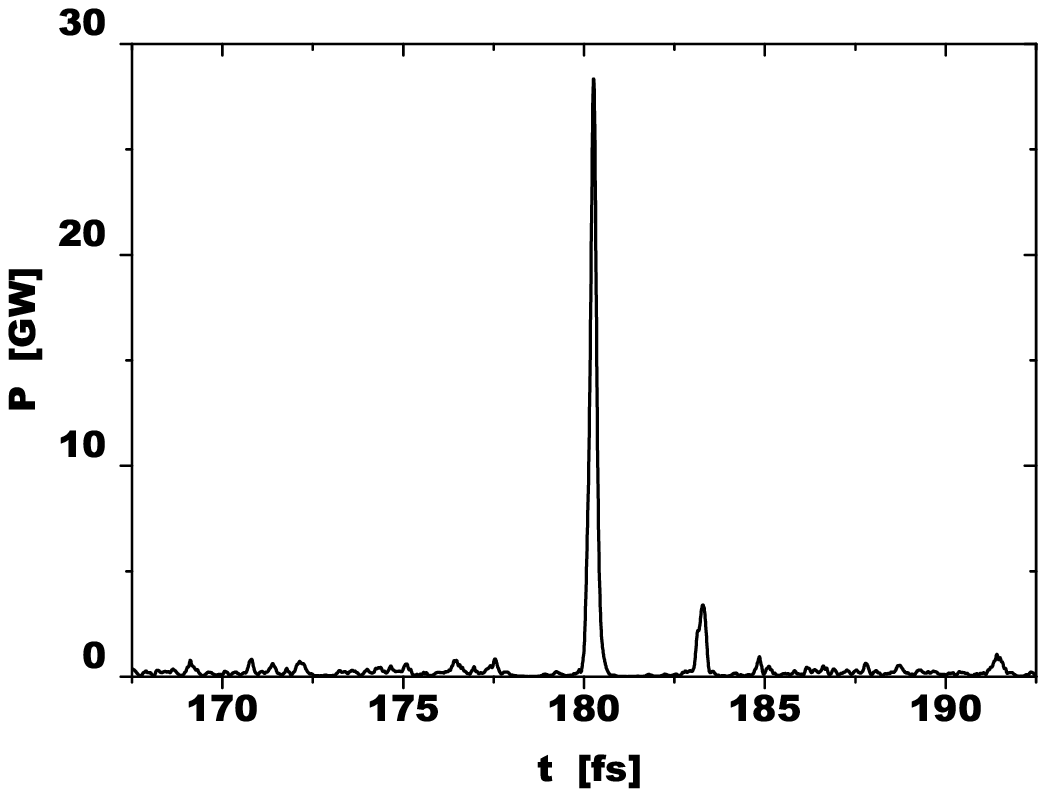,width=0.5\textwidth}

\vspace*{-61mm}

\hspace*{0.5\textwidth}
\epsfig{file=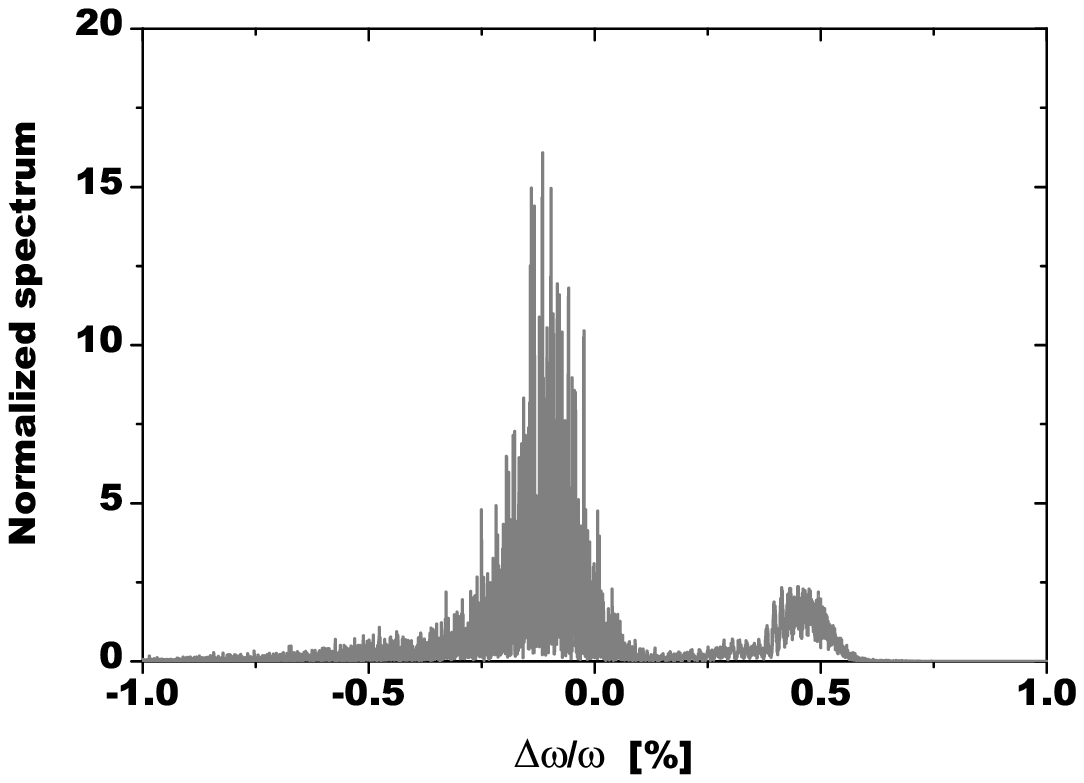,width=0.5\textwidth}

\epsfig{file=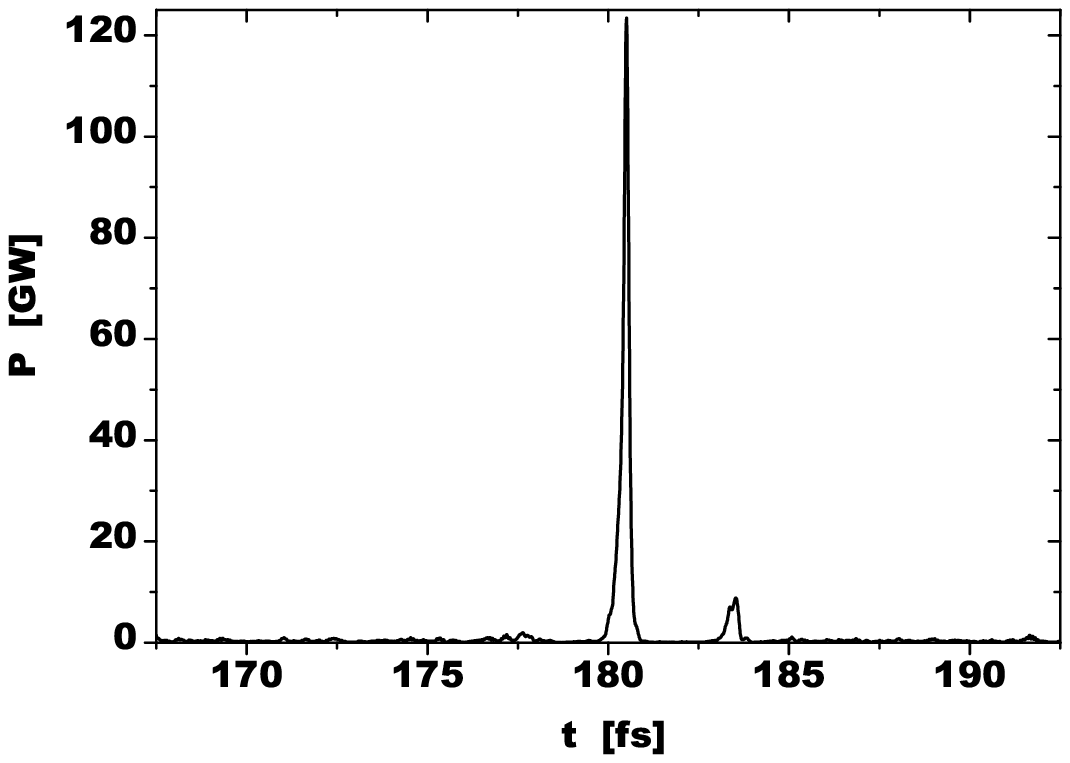,width=0.5\textwidth}

\vspace*{-61mm}

\hspace*{0.5\textwidth}
\epsfig{file=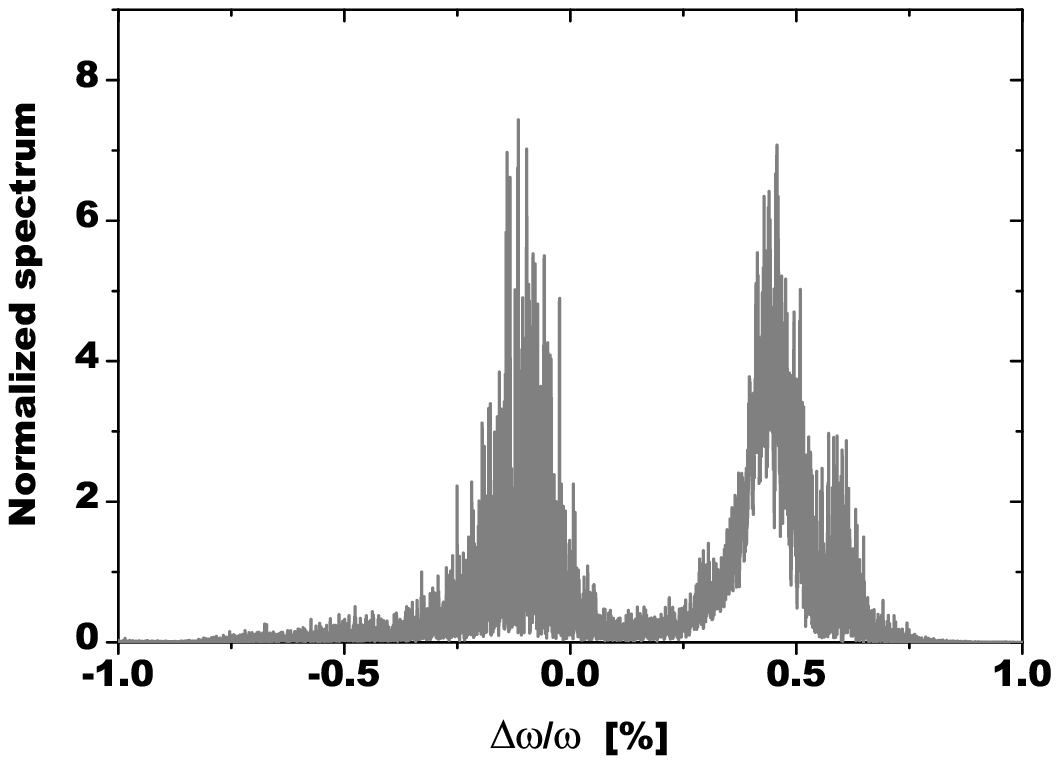,width=0.5\textwidth}

\caption{
Temporal (left column) and spectral (right column)
evolution of the radiation pulse along the undulator.
Upper, middle, and lower plots correspond to the undulator
lengths of 57, 85, and 100~m.
Dashed line shows energy modulation of the electron bunch
}
\label{fig:ps2023}
\end{figure}

\begin{figure}[tb]

\begin{center}

\epsfig{file=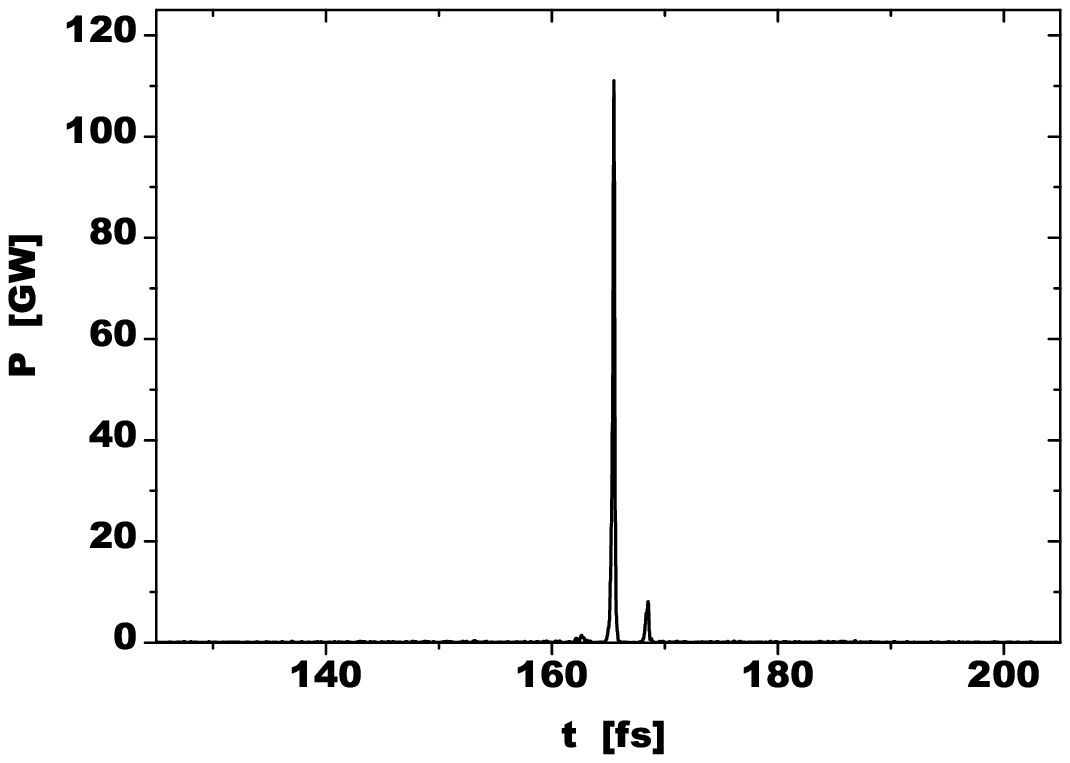,width=0.7\textwidth}

\end{center}

\epsfig{file=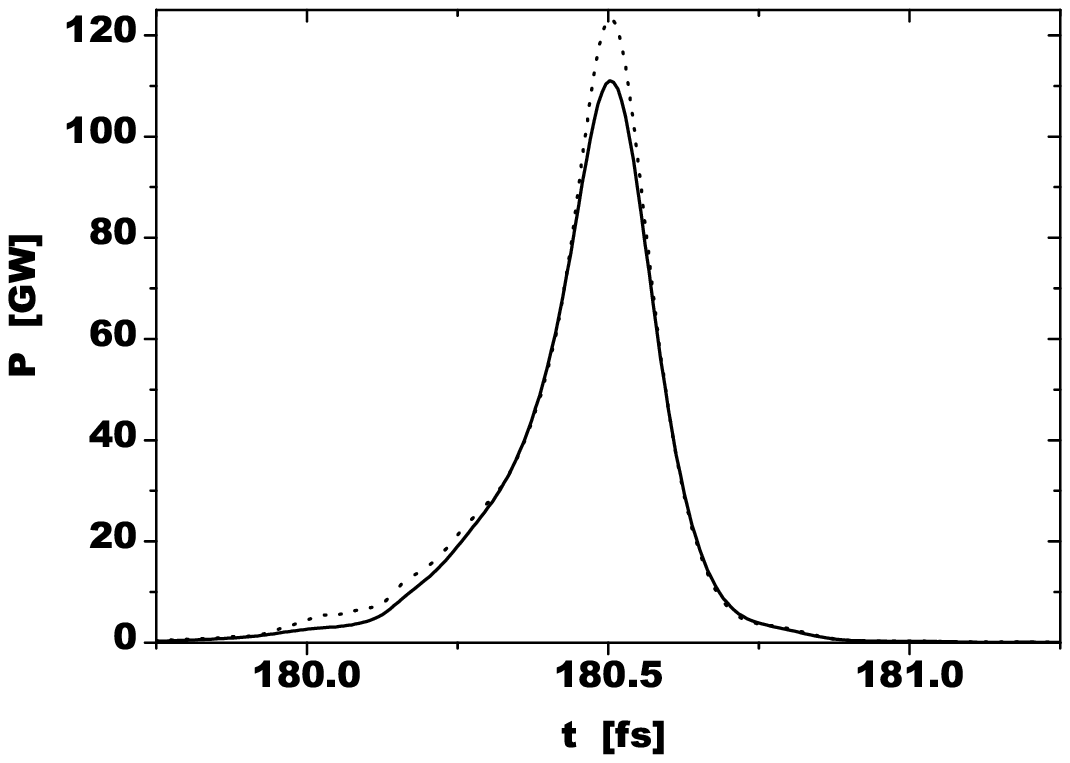,width=0.5\textwidth}

\vspace*{-61mm}

\hspace*{0.5\textwidth}
\epsfig{file=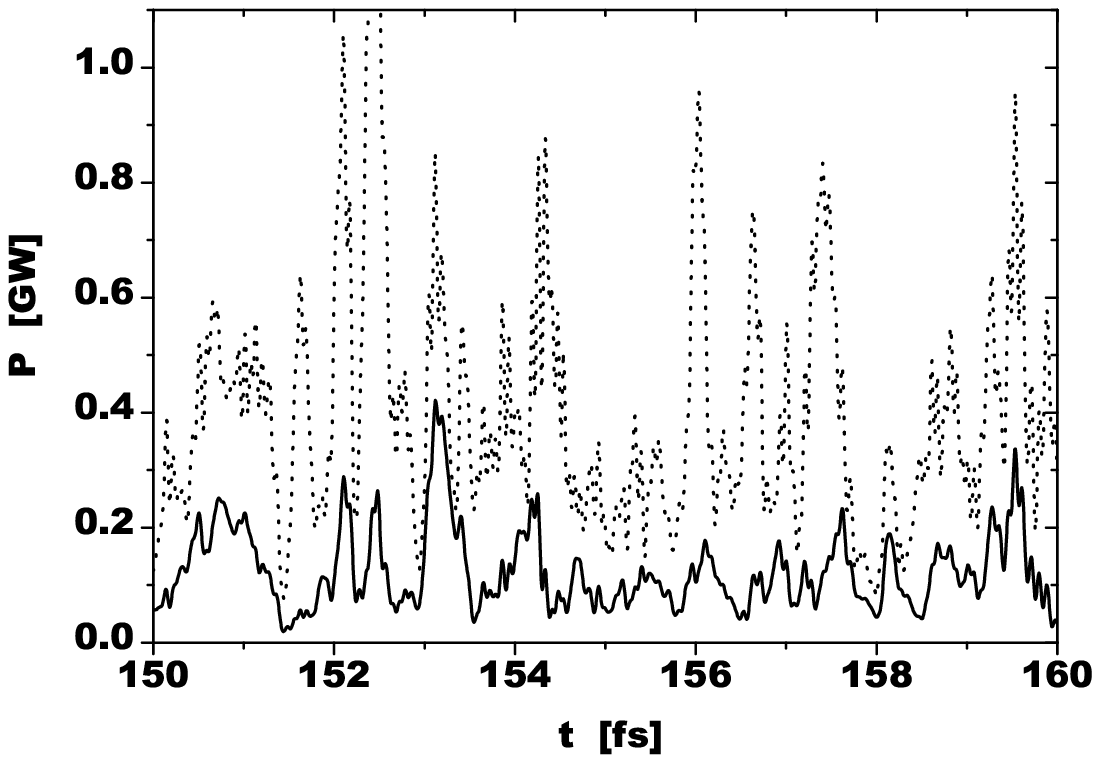,width=0.5\textwidth}

\caption{
Temporal structure of the radiation pulse after monochromator with
0.5\% linewidth tuned to the frequency of the main maximum.
Undulator
length is 100~m. Plots at bottom show enlarged view of the top plot.
Dotted lines show temporal profile of the radiation pulse before
monochromator
}
\label{fig:m2334}
\end{figure}

\begin{figure}[tb]

\begin{center}

\epsfig{file=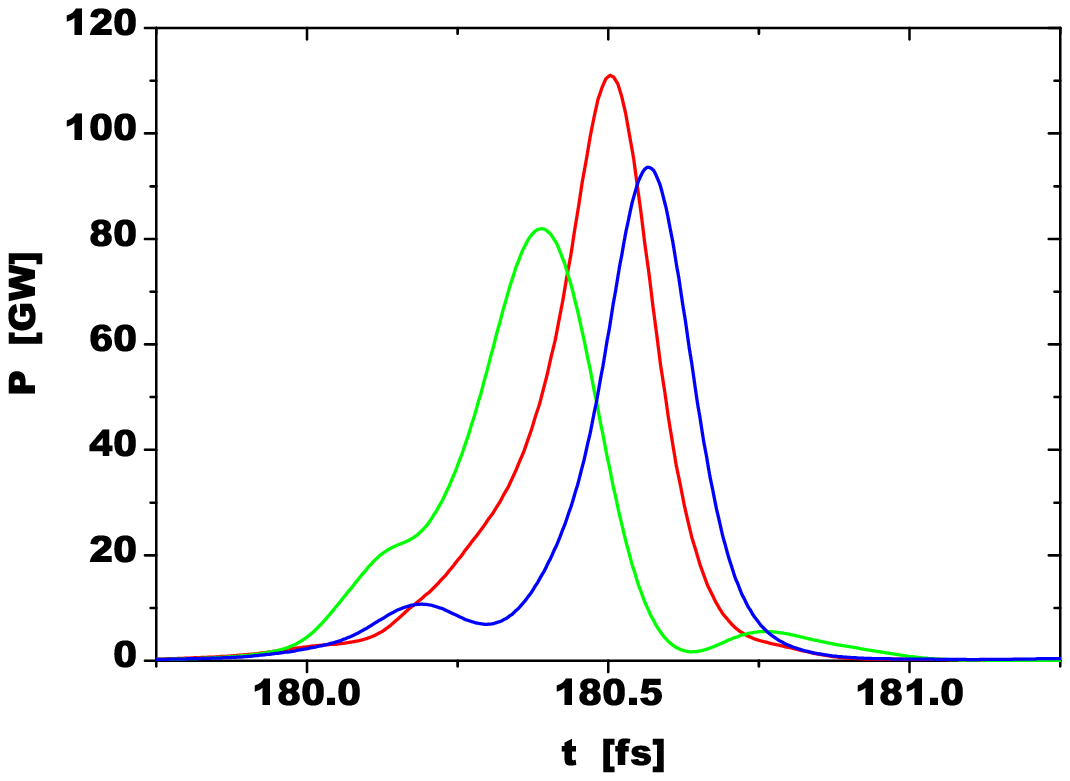,width=0.7\textwidth}

\end{center}

\caption{
Shot-to-shot fluctuations of the
radiation pulse after monochromator with
0.5\% linewidth tuned to the frequency of the main maximum.
Undulator
length is 100~m
}
\label{fig:mpeak}
\end{figure}

Figure~\ref{fig:pz2023} shows evolution of energy in the radiation
pulse along the undulator. Generation of powerful attosecond X-ray
pulses is performed in three-step procedure. Electron beam enters the
first undulator section. Amplification process develops from the shot
noise in the electron beam. Figure~\ref{fig:m2318} shows temporal
structure of the radiation pulse in the end of the first undulator
section. Top plots in Fig.~\ref{fig:ps2023} show enlarged view of
temporal structure and spectral structure. A signature of the slice
beam modulation is clearly seen in the temporal structure of the
radiation pulse. The dotted line in this figure shows the initial
energy modulation of the electron beam. The temporal structure of the
radiation pulse has a clear physical explanation. The FEL process is
strongly suppressed in the regions of the electron bunch with large
energy chirp, and only regions of the electron bunch with nearly zero
energy chirp produce radiation. From a physical point of view each part
of the bunch near the local extremum of the energy modulation can be
considered as an isolated electron bunch of short duration. At the
chosen parameters of the system its duration is about 300~attosecond
which is about of coherence time. Thus, it is not surprising that only
a single radiation spike is produced by each area near the local
extremum. Note that each of these spikes (wavepackets) has significant
frequency offset from the frequency of the main radiation pulse. For
instance, frequency detuning of the spike corresponding to time $t =
165$~fs is about 0.5\%. It is hardly noticeable on the scale of
complete spectrum of the radiation pulse, since its contribution to the
total radiation energy is a fraction of a per cent.

At the next step the electron beam passes the dispersion section. The
dispersion section performs two actions. First, it provides delay of
the electron bunch by 15~fs with respect to the radiation pulse, such
that the radiation produced by modulated slice of the electron bunch
slips forward to "fresh", nonmodulated part of the bunch. Second, the
strength of dispersion section is sufficient for suppression of the
beam bunching. As a result, the amplification process in the second
undulator section starts with "fresh" electron beam and radiation seed
produced by the first undulator section. The undulator of the second
section is tuned to $\Delta \omega/\omega = 0.6$\% in order to provide
the resonance with the radiation produced by the slice of the electron
bunch having maximum energy offset. Under such conditions only a single
spike is amplified effectively as it is illustrated with plots in
Fig.~\ref{fig:ps2023}. The rest part of the seed radiation pulse has
significant detuning and propagates without visible interaction with
the electron bunch. Amplitude of a single attosecond pulse grows
exponentially with the undulator length, and enters to saturation
regime with peak power of about 20~GW. In order to increase
peak power to 100~GW level final (the third) undulator section is
tapered. Bottom plots in Fig.~\ref{fig:ps2023} illustrate typical
temporal and spectral structure of the radiation pulse at the exit of
the SASE undulator. In principle, much higher than 100~GW peak power in
the attosecond pulse can be achieved. However, one should keep in mind
the background produced by the rest of the electron bunch.
The seed radiation from the first undulator section is not a problem,
it has large frequency offset and can be simply filtered by multilayer
monochromator. The main limiting factor is the growth of the radiation
from the shot noise which has the same frequency spectrum as attosecond
pulse. When attosecond pulse enters nonlinear stage of amplification,
noise background still continues to grow exponentially, and can reach
high power. This effect imposes a limit on the allowable undulator
length (and on peak power of the attosecond pulse).
Figure~\ref{fig:m2334} shows temporal characteristics of the radiation
pulse at the exit of the undulator (dotted lines). Solid lines in this
plot show properties of the radiation pulse after monochromator with
0.5\% linewidth. It is seen that the contrast of the radiation pulse is
pretty high. Figure~\ref{fig:mpeak} illustrates shot-to-shot
fluctuations of the radiation pulse.

\section{Pump-probe experiments with attosecond SASE FEL}

The proposed high power, ultra-fast X-ray source holds a great promise
as a source of X-ray radiation for such applications as pump-probe
experiments. The time resolution of pump-probe experiment is obviously
determined by the duration of the pump as well as the resolution of the
probing. The pump pulse must always be as short as the desired time
resolution. In typical scheme of a pump-probe experiment the short
probe pulse follows the pump pulse at some specified delay. The signal
recorded by the slow detector then reflects the state of the sample
during the brief probing. The experiment must be repeated many times
with different delays in order to reconstruct the dynamical process.

\begin{figure}[tb]
\begin{center}
\epsfig{file=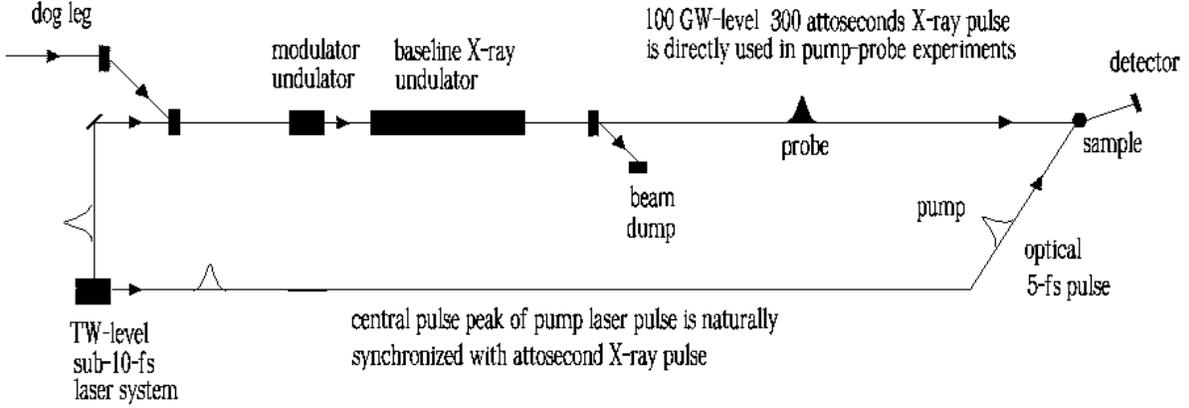,width=1\textwidth}
\end{center}
\caption{Scheme for femtosecond resolution pump-probe
experiments based on the
generation of the 100 GW-level attosecond X-ray pulses directly from
X-ray SASE FEL
}
\label{fig:x6}
\end{figure}

We suggest to combine attosecond X-ray pulses with fs optical pulses
generated in the seed Ti:sapphire laser system (see Fig.  \ref{fig:x6})
for pump-probe experiments. In this case attosecond X-ray pulse is
naturally synchronized with its fs optical pulse, and time jitter is
cancelled.  Another advantage of the proposed scheme is the possibility
to remove all X-ray optical elements between the X-ray source and a
sample and thus to directly use the probe attosecond X-ray pulse.
Usual optical elements are used for seed laser beam splitting and
tunable delay.  It should be possible to achieve a timing accuracy
close to duration of the half period of the seed  laser pulse (1 fs),
allowing an unprecedented insight into the dynamics of electronic
excitations, chemical reactions, and phase transitions of matter, from
atoms, through organic and inorganic molecules, to surface, solids and
plasma.

Let us discuss in more detail an attosecond pulse
contrast in the proposed scheme for pump-probe experiments based on the
use the direct X-ray beam. We wish to consider one example which shows
the relationship between the signal and background that is easy to
understand.  We choose for emphasis here experiments aimed at
measuring the lifetime of electronic excitation. First, the time
dependence of electronic occupation is assumed to be known -- this is
an exponential dependence $n(t) = n_{0}\exp(-t/\tau_{0})$. The typical
lifetime of electronic excitation is in the femtosecond range. A
time-resolved experiment inherently begins with initiation of the
process under study (electronic excitation) at some more or less
accurately defined instant in time. The number of
photocounts detected during attosecond pulse  (probe) is directly
proportional to the physical parameter (electronic occupation number)
to be measured: $K =
AI_{\mathrm{a}}\tau_{\mathrm{a}}n(\tau)$, where
$I_{\mathrm{a}}$ is the intensity of attosecond pulse,
$\tau_{\mathrm{a}}$ is the attosecond pulse duration, and $\tau$ is the
delay between pump and probe pulses. On the other hand, the number of
background photocounts is  proportional to the integrated value of the
physical parameter:  $K =
A\int\limits^{\infty}_{0}I_{\mathrm{b}}(t)n_{0}\exp(-t/\tau_{0})d t$,
where $I_{\mathrm{b}}(t)$ is instantaneous background intensity of SASE
radiation. At time $t = 0$ the pump pulse perturbs the sample. As a
result, the average number of background photocounts is about $\langle
K\rangle \simeq A\langle I_{\mathrm{b}}\rangle\tau_{0}n_{0}$. The ratio
of attosecond intensity and background intensity is about
$I_{\mathrm{a}}/I_{\mathrm{b}} \simeq 400$.  On the other hand, the
ratio of attosecond pulse duration and characteristic time of
(sub-10-fs) physical process is about $\tau_{\mathrm{a}}/\tau_{0}
\simeq 0.1$ only.  Thus, we find that effects of SASE background
radiation are not important in experiments for the study of the
dynamics of the sub-10-fs processes.

It is obvious that proposed pump-probe techniques could be applied at
longer time-scale, too. In particular, there exists a wide interest in
the extension of light-triggered time-resolved studies to the sub-100-fs
timescale. Now we return to the question about an attosecond pulse
contrast. According to our discussion above, the number of background
photocounts in this case is proportional to integral over the whole SASE
radiation pulse with duration of about 300 times longer than the
attosecond pulse.  Calculation shows that in this case the ratio of
attosecond pulse energy to SASE radiation pulse energy reaches a value
of about 1:1. The number of photocounts as a function of delay time has
thus a peak to background of 2 to 1.  These sub-100-fs studies we can
refer as time-resolved experiments with the background, as opposed to the
background-free time-resolved experiments for study of the sub-10-fs
processes. It should be noted that the final time resolution of
pump-probe experiments with the background is not worse than that
of background-free experiments with the same pump
and probe pulse duration, but we secure this resolution with a much
higher number of independent measurements.

A scheme of pump-probe experiments for the study of the dynamics of
the sub-100-fs processes can be modified to increase significantly the
contrast of output attosecond X-ray pulses.  A reliable method to
decrease the background might be to use a broadband monochromator. One
can align the monochromator so that the peak of the SASE radiation
spectrum at reference frequency is blocked.  To reach the required
value of monochromatization is not a problem.  For 0.15 nm wavelength
range, Bragg diffraction is the main tool used for such purposes. In
this case, one has to care that the short pulse duration is preserved.
We are discussing here only multilayer X-ray mirrors, which have the
largest relative bandwidth
\cite{ml1,ml2,ml3}. Layered synthetic
materials -- multilayers -- are layered structures with usually two
alternating materials: a low and high density materials.  They play an
important role in synchrotron X-ray optics. Typical multilayers used as
optical elements at the third generation synchrotrons provide a spectral
bandwidth of 0.5 to 5\%. Typical glancing angles are of the order of a
degree. As a rule, from 100 to 400 periods participate in effective
reflection in such mirrors. About 90\% peak reflectivity was achieved
for wavelengths around 0.1 nm.  Analysis presented in this paper  shows
that this technique has a potential to increase single-spike contrast
(ratio of attosecond pulse power to background power) up to 2000. This
means that in the proposed scheme of pump-probe experiments with
multilayer monochromator the effects of SASE background radiation are
not important  even in the 100 fs time range.

\section{Conclusion}

Today there are at least two possible attosecond X-ray sources for
light-triggered, time-resolved experiments associated with the X-ray
SASE FEL: the "attosecond X-ray parasitic" \cite{atto-1} and the
"attosecond X-ray dedicated" source mode proposed in this paper. The
simplest way to obtain attosecond X-ray pulses from XFEL is to use
"parasitic" technique which is proposed in \cite{atto-1}. It also would
offer the possibility of providing a beam to a pump-probe experiments
with the XFEL that has a precise, known and tunable time interval
between the laser and X-ray sources. More power of attosecond pulse
could be obtained using the XFEL for dedicated attosecond X-ray pulse
production as described in this paper.

Both types of attosecond X-ray sources are important, their roles are
complementary, and one type can not replace the other. The baseline
gap-tunable XFEL undulator offers simplicity and flexibility. Different
operational modes can be realized with the undulator control system.
The baseline XFEL operates in uniform (maximum gain) mode. With no
constraints on baseline FEL operation, technique for production of
attosecond X-ray pulses which is proposed in \cite{atto-1} could be
used.  If dedicated attosecond beamtime will be available, an undulator
could be tapered (and magnetic chicane could be switched on) to provide
more power in attosecond pulses.

\section*{Acknowledgments}

We thank R.~Brinkmann, G.~Gr\"ubel,
J.R.~Schneider, A.~Schwarz, and
D.~Trines for interest in this work.


\begin{thebibliography}{99}

\bibitem{laser-atto1}
P.~Paul,
Science 292(2001)1689.

\bibitem{laser-atto2}
M.~Hentchel et al.,
Nature 414(2001)509.

\bibitem{xfel-tdr}
P.~Audebert et al., ``TESLA XFEL: First stage of the X-ray laser
laboratory -- Technical design report (R.~Brinkmann et al., Eds.)'',
Preprint DESY 2002-167.

\bibitem{lcls-cdr}
The LCLS Design Study Group, LCLS Design Study Report, SLAC reports
SLAC- R521, Stanford,  1998.

\bibitem{atto-1}
E.L. Saldin, E.A. Schneidmiller and M.V. Yurkov,
Preprint DESY 04-013, 2004. Submitted to Optics Communications.

\bibitem{book}
E. L. Saldin, E. A. Schneidmiller and M. V. Yurkov,
The Physics of Free Electron Lasers (Springer-Verlag, Berlin) 1999.

\bibitem{yu-freshb}
L. H. Yu and I. Ben-Zvi,
Nucl. Instrum. and Methods A393(1997)96.

\bibitem{fast}
E. L. Saldin, E. A. Schneidmiller and M. V. Yurkov,
Nucl. Instrum. and Methods {\bf A 429}(1999)233.

\bibitem{ml1}
P.~Deschamps et al.,
J. Synchrotron Rad. 2(1995)124.

\bibitem{ml2}
D.~Windt,
Appl. Phys. Lett. 74(1999)2890.

\bibitem{ml3}
Ch. Morawe et al.,
SPIE Proc. 4145(2000)61.

\end{thebibliography}
\end{document}